\documentclass[]{aa}
\usepackage{graphicx}
\usepackage{color}
\usepackage{natbib}
\begin{document}

\title{Evidence from stellar rotation for early disc dispersal owing to close companions}
\author{S.\,Messina\inst{1}}
\offprints{Sergio Messina}
\institute{INAF-Catania Astrophysical Observatory, via S.Sofia, 78 I-95123 Catania, Italy \\
\email{sergio.messina@inaf.it}
}

\date{}
\titlerunning{Evidence for early rotation spinning up}
\authorrunning{S.\,Messina et al.}
\abstract {Young ($\la$ 600\,Myr) low-mass stars (M $\la$1M$_{\odot}$)  of equal mass exhibit a distribution of rotation periods. At the very early phases of stellar evolution, this distribution is set by the star--disc locking mechanism, 
which forces stars to rotate at the same rate as the inner edge  of the disc. The primordial disc lifetime and consequently the duration of the disc-locking mechanism, can be significantly shortened by the presence of a close companion, making the rotation period distribution of  close binaries different from that of either single stars or wide binaries. }{We use new data to investigate and better constrain the range of ages, the components separation, and the  mass ratio  dependence at which the rotation period distribution has been significantly affected by the disc dispersal that is enhanced by close companions.} {We select a sample of close binaries in the Upper Scorpius association (age $\sim$8\,Myr) whose components have  measured the separation and the rotation periods and compare their period distribution with that of coeval stars that are  single stars.} {We find that components of close binaries have, on average, rotation periods that are shorter  than those of single stars.  More precisely, binaries with approximately equal-mass components (0.9 $\le$ M2/M1 $\le$ 1.0) have rotation periods that are shorter than those of  single stars  by $\sim$0.4\,d  
on average; the primary and secondary components of binaries with smaller mass ratios (0.8 $<$ M2/M1 $<$ 0.9) have rotation periods that are shorter than those of single stars by $\sim$1.9\,d  and  $\sim$1.0\,d  on average, respectively.  A comparison with the older 25-Myr $\beta$ Pictoris association shows that whereas in the latter, all close binaries with projected separation $\rho$ $\le$ 80\,AU rotate faster than single stars, in the Upper Scorpius this is only the case for about 70\% of stars.} {We interpret the enhanced rotation in close binaries with respect to single stars as the consequence of an early disc dispersal induced by the presence of close companions. The enhanced rotation suggests that disc dispersal timescales are longest for single stars and shorter for close binaries.}
\maketitle
\rm

\section{Introduction}

Late-type stars (M $\la$ 1\,M$_{\odot}$) with similar mass and age show a distribution of rotation periods. The width of this distribution decreases as stars age, until a one-to-one correspondence between mass and rotation period is reached by an age of about 0.6\,Gyr (e.g. \citealt{Delorme11}). Such a distribution is thought to arise from a distribution of the initial rotation periods, that is the rotation periods set during the disc-locking phase. Indeed, at the early stages of their life, most if not all stars are characterised  by the presence of a primordial circumstellar disc that, while accreting mass and transferring angular momentum onto the star, leaves its imprinting, that is it fixes the value of the initial stellar rotation period. This happens by means of the disc-locking mechanism, which forces the outer layers of the star to rotate for a few million years at the rotation rate of the primordial disc inner edge \citep{shu94}. 
After the disc dispersal, this imprinting remains for a long time, until the one-to-one correspondence is reached between mass and period and all memory of the initial rotation period becomes lost.\\
\indent
Another parameter that effectively contributes to the observed distribution of rotation periods among coeval equal-mass stars is the disc lifetime. The primordial disc lifetime is generally not longer than about 10\,Myr (\citealt{Ingleby14}; \citealt{Ribas14}), but there are exceptions (see, e.g. \citealt{Frasca15}). However, this lifetime is variable and can be significantly shortened by different factors, such as the gravitational perturbance effects by a close companion. Once a star experiences either an early disc dispersion or inner disc truncation, its rotation rate starts spinning-up earlier than equal-mass disc-bearing stars, because of the radius contraction and angular momentum conservation, gaining a shorter rotation period in comparison.\\
\indent
Indeed, evidence has been accumulated showing that among coeval stars (i.e. members of the same association or cluster) members without discs tend to rotate faster than those with discs (\citealt{Kraus16}, \citealt{Cieza09}). Furthermore, components of close binaries tend to have a smaller occurrence of discs and to exhibit shorter rotation periods (\citealt{Stauffer16}, \citeyear{Stauffer18}; \citealt{Rebull18}).\\
\indent
We intend to use the rotation period as a diagnostic to explore the effective existence of a disc dispersal enhancement, and its dependence on the separation between the binary components, on their mass ratio, and on age. As already mentioned, the process of enhanced disc dispersal takes place before 10 Myr of age. Therefore, our investigation focuses on clusters and associations in this range of ages. Nonetheless, valuable information can also be derived from the analysis of older associations. Indeed, in  the 25-Myr  \object{beta Pic Association}, we found clear evidence for earlier disc dispersal induced by the presence of close companions. At that age, members of close binaries (projected separation $<$ 80 AU) all rotate faster than their single counterparts \citep{Messina17}. \\
\indent
Recently, two relevant investigations by \cite{Rebull18} and \cite{Tokovinin18} made accurate rotation period measurements available and newly imaged and spatially resolved a number of close binaries in the \object{Upper Scorpius} association at an age of $\sim$8 Myr. This new information enables us to push our investigation of the effects of binarity on disc dispersal, and then on rotation,  back to a much younger age.\\

In this paper, we report the results of our analysis of the dependence of rotation on binarity at an age of about 8 Myr among the low-mass candidate members of the young stellar association Upper Scorpius (USco). In Sect.\,2 we describe the sample selection and the data. In Sect.\,3, we present our analysis and in Sect.\,4 we present a discussion and our interpretation. Conclusions are given in Sect.\,5.

\begin{figure}
\begin{minipage}{10cm}
\includegraphics[scale = 0.32, trim = 0 0 0 0, clip, angle=90]{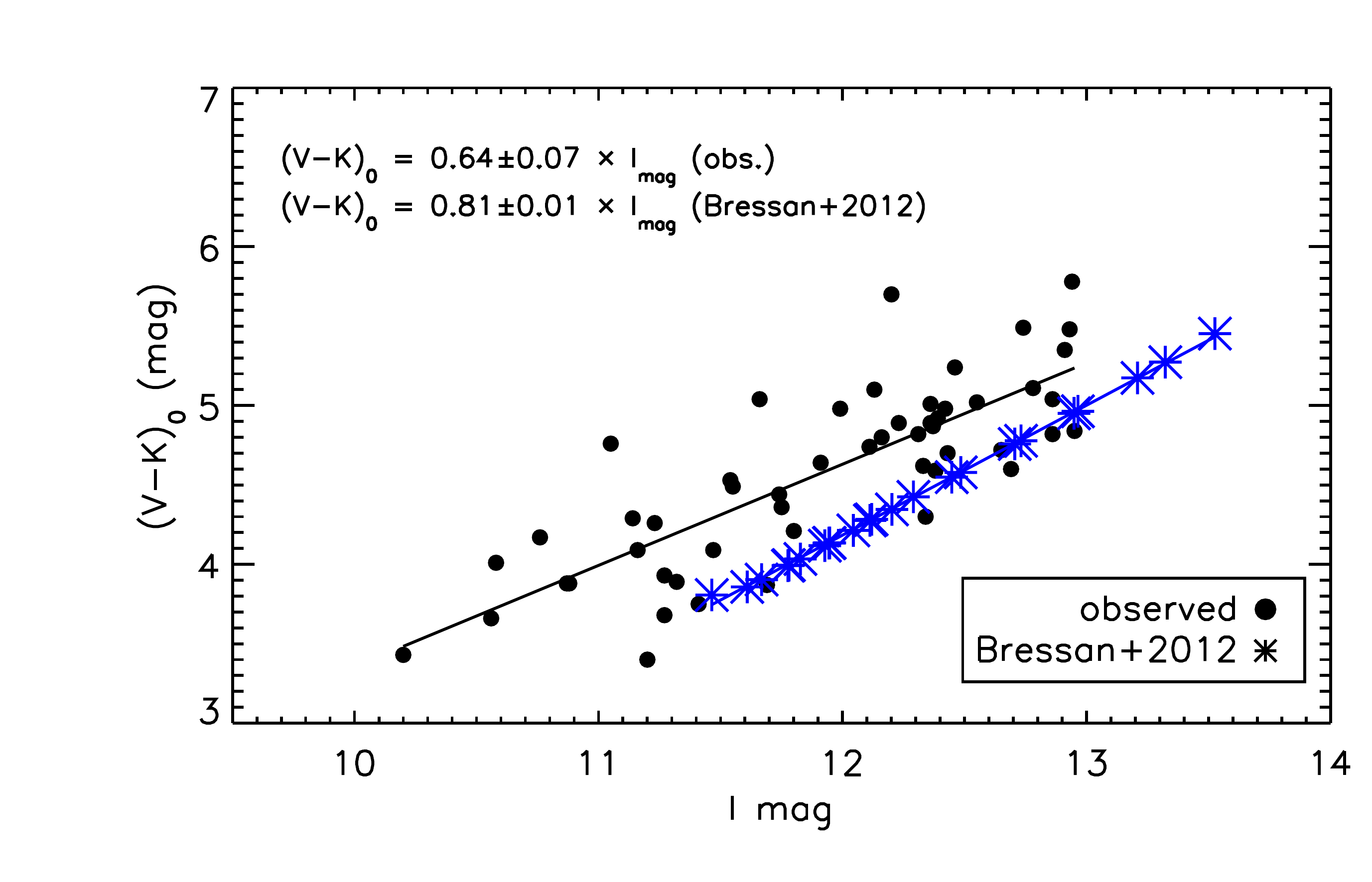}
\end{minipage}
\caption{\label{VK-mag} Colour de-reddened (V$-$K)$_0$ vs. observed I mag (bullets) for the sample observed by \citet{Tokovinin18}.  Blue asterisks represent the model I magnitudes corrected for the distance modulus 5.8\,mag (\citealt{Wilkinson18}), and the model colours from \citet{Bressan12}. Solid lines are linear fits.
We note that the model values are displaced on average by 0.75\,mag from the magnitude of observed binary systems.}
\end{figure}

\begin{figure*}
\begin{minipage}{18cm}
\centering
\includegraphics[scale = 0.5, trim = 0 0 0 0, clip, angle=90]{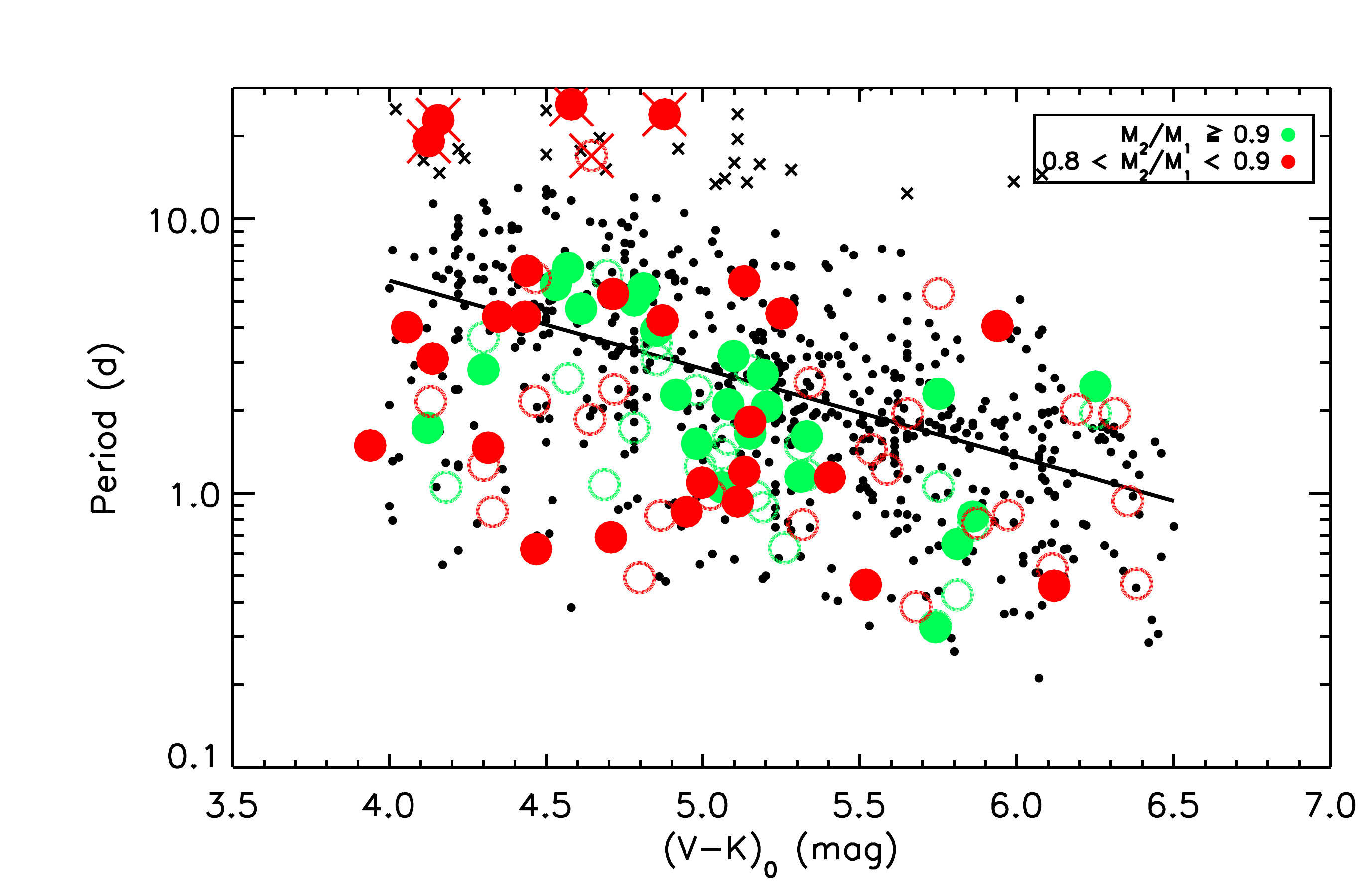}
\end{minipage}
\caption{\label{color-period} Distribution of stellar rotation periods vs. dereddened colour for candidate members of the Upper Scorpius association from \citet{Rebull18} in the M0--M6 spectral range. Small dots are the rotation periods of single stars as listed in Table 1 of  \citet{Rebull18}. Filled and open bullets are periods P1 and P2 of components of close binaries, respectively. Components of binary systems whose mass ratio is M$_2$/M$_1$ $\ge$ 0.9   (respectively 0.8 $\le$
M$_2$/M$_1$ $\le$ 0.9)\ are shown in green (respectively red). The solid line is a  fit to the rotation periods of single stars (see Eq.\,(5)). Crossed symbols are outliers excluded from the fit computation and the following analysis.}
\end{figure*}

\section{Sample selection}
\citet{Rebull18} recently measured numerous rotation periods ($\sim$1000) in a sample of about 1300 candidate members of the young association USco. Interestingly, they report the finding of a sample of 239 candidate members showing multi-periodic light variations. They inferred periodicities by means of Lomb-Scargle periodogram analysis \citep{Scargle82} of the photometric time series collected during the Kepler K2 
campaign.\\
\citet{Tokovinin18} observed  by means of spickle interferometry 129 of the brighter stars ($I < 13$\,mag and 3 $<$ (V$-$K)$_0$ $<$ 6\,mag) of the multi-periodic sample of stars found by \citet{Rebull18}. They  probed the presence of companions in the separation range from 0.04$^{\prime\prime}$ to $\sim$3$^{\prime\prime}$, corresponding to separations from $\sim$5 to  $\sim$400\,AU. As a result of their investigation, they spatially resolved 70 of them, giving additional support to the interpretation that multi-periodic stars are mostly binary stars. 
The sample selection criterion adopted by  \citet{Tokovinin18} favoured the detection of binaries whose components have comparable flux; indeed most resolved components have a magnitude difference  $\Delta I$ $<$ 1\,mag.

Compared to an isochrone of 8\,Myr \citep{Bressan12}, \citet{Tokovinin18} found their sample of resolved binaries to be displaced on average by $\simeq$+0.75\,mag above the isochrone, as expected for binary stars with nearly equal-mass components. They measured  $I_C$ magnitude,  angular separation, and magnitude difference between the resolved components, and derived the mass of the primary component (see their Table 1). 

From the original Tokovinin sample of 70 resolved binaries, we selected a subsample of 49 targets in the colour range 4 $<$ (V$-$K)$_0$ $<$ 6.5\,mag in order to focus on M-type stars. 

The close binaries in our sample have mass ratios in the range 0.8 $\le$ M$_2$/M$_1$ $\le$ 1, with only one binary with M$_2$/M$_1$ = 0.65. The mass ratio is derived using the mass of the primary component derived by \citet{Tokovinin18} and the mass-$I$$_{\rm mag}$ relation from the 8-Myr isochrone \citep{Bressan12}, transforming the observed magnitude difference $\Delta I$ into mass difference $\Delta$$M$ 

\begin{equation}
\Delta M = M_1 - M_2 = \Delta I \times 0.197\pm 0.014
,\end{equation}
\noindent
where 0.197 is the slope of the   mass-$I$$_{\rm mag}$ relation in the M0--M6 spectral range.

\section{Analysis}
In the spectral range from M0 to about M6, we investigate if any difference exists between the period distribution of  the  49 close binaries with known component separation that we selected from the \citet{Tokovinin18} sample and the period distribution of all single-star candidate members taken from \citet{Rebull18}.\\
\indent
 \citet{Rebull18} assumed that the periodicity P1 (which corresponds to the highest power peak in the Lomb-Scargle periodogram) is the rotation period of a single star or of the primary component of a multiple system. The second periodicity P2 (which corresponds to the second power peak
in the Lomb-Scargle periodogram in order of decreasing power), when detected,  is the rotation period of the secondary component of a multiple system. Therefore, P1 and P2 are the rotation periods of the components of  the resolved close binaries in our analysis.\\

The dereddened (V$-$K)$_0$ colours provided by \citet{Rebull18} for each photometric binary refer to the integrated system, whereas the primary and the secondary components have colours that are bluer and redder, respectively, than the integrated color. To derive the appropriate colours for both components we proceeded as follows. First, we used the \citet{Bressan12} models for the age of 8\,Myr to derive the colour correction $\Delta(V-K)_{0P}$ for the primary component, 
\begin{equation}
\Delta(V-K)_{0P} = -2.5\log\left(\frac{1+ \frac{F_{K2}}{F_{K1}}}{1+ \frac{F_{V2}}{F_{V1}}}\right)
,\end{equation}
where F$_{V1}$ and F$_{V2}$ are the integrated fluxes in the V band, F$_{K1}$ and F$_{K2}$ are those in the K band, and P stands for primary component. 
We adopted the colours (V$-$K)$_{0P}$ = (V$-$K)$_0$ $+$ $\Delta$(V$-$K)$_{0P}$ for the primary components.\\ 
Similarly, for the secondary
component, we computed the colour correction $\Delta(V-K)_{0S}$ ,
\begin{equation}
\Delta(V-K)_{0S} = -2.5\log\left(\frac{1+ \frac{F_{K1}}{F_{K2}}}{1+ \frac{F_{V1}}{F_{V2}}}\right)
,\end{equation}
and adopted the colors (V$-$K)$_{0S}$ = (V$-$K)$_0$ $+$ $\Delta$(V$-$K)$_{0S}$ .\\ 

The colour correction $\Delta(V-K)_{0S}$ for the secondary component (S), can also be computed by using the magnitude difference between the components $\Delta I$  measured by \citet{Tokovinin18} and the  linear regression  coefficient (a1 = 0.64$\pm$0.07) between the observed $I$ magnitude provided by \citet{Tokovinin18} and the reddening-corrected colour (V$-$K)$_0$, as shown in Fig.\,\ref{VK-mag}: 
\begin{equation}
\Delta(V-K)_{0S} = 0.64 \pm 0.07 \times \Delta I
.\end{equation}

We note that the model values of magnitude and colour from \citet{Bressan12} yield a larger value for the coefficient (a2 = 0.81$\pm$0.07).  We find that the use of a1 and a2 produce colours for the secondary components that are 0.08\,mag and 0.15\,mag redder,
respectively, than those computed from Eq.\,(3). We found that the choice of method used to compute the colours of the secondary components has no effect on the results of the following analysis.

The relevant quantities taken from \citet{Tokovinin18} and from \citet{Rebull18} and the new ones computed in the present study for the selected 49 targets are listed in Table\,\ref{Table1}.\\
\rm
In Fig. \ref{color-period} we plot the rotation period P versus (V$-$K)$_0$ colour of all single-period candidate members and overplot the rotation period P1  and P2  of the components of the 49 close binaries versus their colours, as computed according to Eqs. (2) and (3).     All stars whose period residuals were larger than 3$\sigma$ (i.e. P - P$_{fit}$ $>$ 9.2\,d) were rejected (crosses in Fig.\,\ref{color-period}), and the new fit was computed:
\begin{equation}
\log _{10} P = -0.304\pm0.020\times(V-K)_0 + 1.94\pm0.10
,\end{equation}
where P is the rotation period in days.\\
\indent
Before proceeding with our analysis, we must consider what follows.  
In unresolved  close binaries, a fainter component whose flux ratio is for example F$_2$/F$_1$ $\ge$ 0.6 can exhibit activity-induced flux variability with amplitude $\ge$ 75\% of that exhibited by the primary component. This means that, depending on the specific properties of the activity patterns on the photosphere of both components, the variability arising from the secondary component may be dominant and produce the most powerful peak in the periodogram. In this circumstance the primary period P1 should rather be attributed to the secondary component. Therefore, in our analysis we consider that rotation periods P1 and P2  assigned by \citet{Rebull18} to the primary and secondary components, respectively, in the case of close binaries with F$_2$/F$_1$ $\ge$ 0.6, which corresponds to $\Delta I$ $\le$ 0.5\,mag ($\Delta$M $\le$ 0.1\,M$_{\odot}$), may be exchanged.
\begin{figure}
\begin{minipage}{10cm}
\includegraphics[scale = 0.3, trim = 0 0 0 0, clip, angle=90]{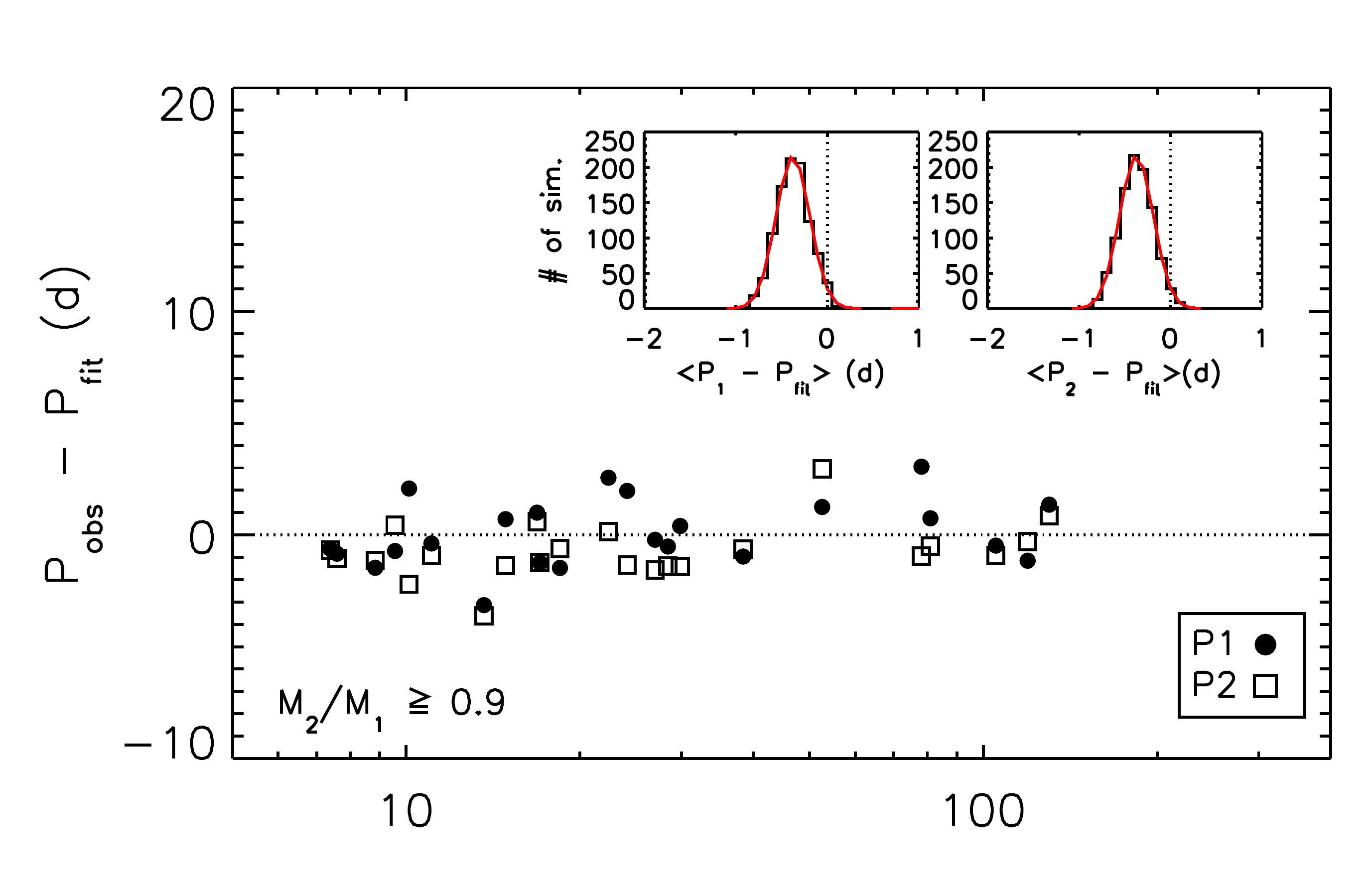}
\includegraphics[scale = 0.3, trim = 0 0 0 0, clip, angle=90]{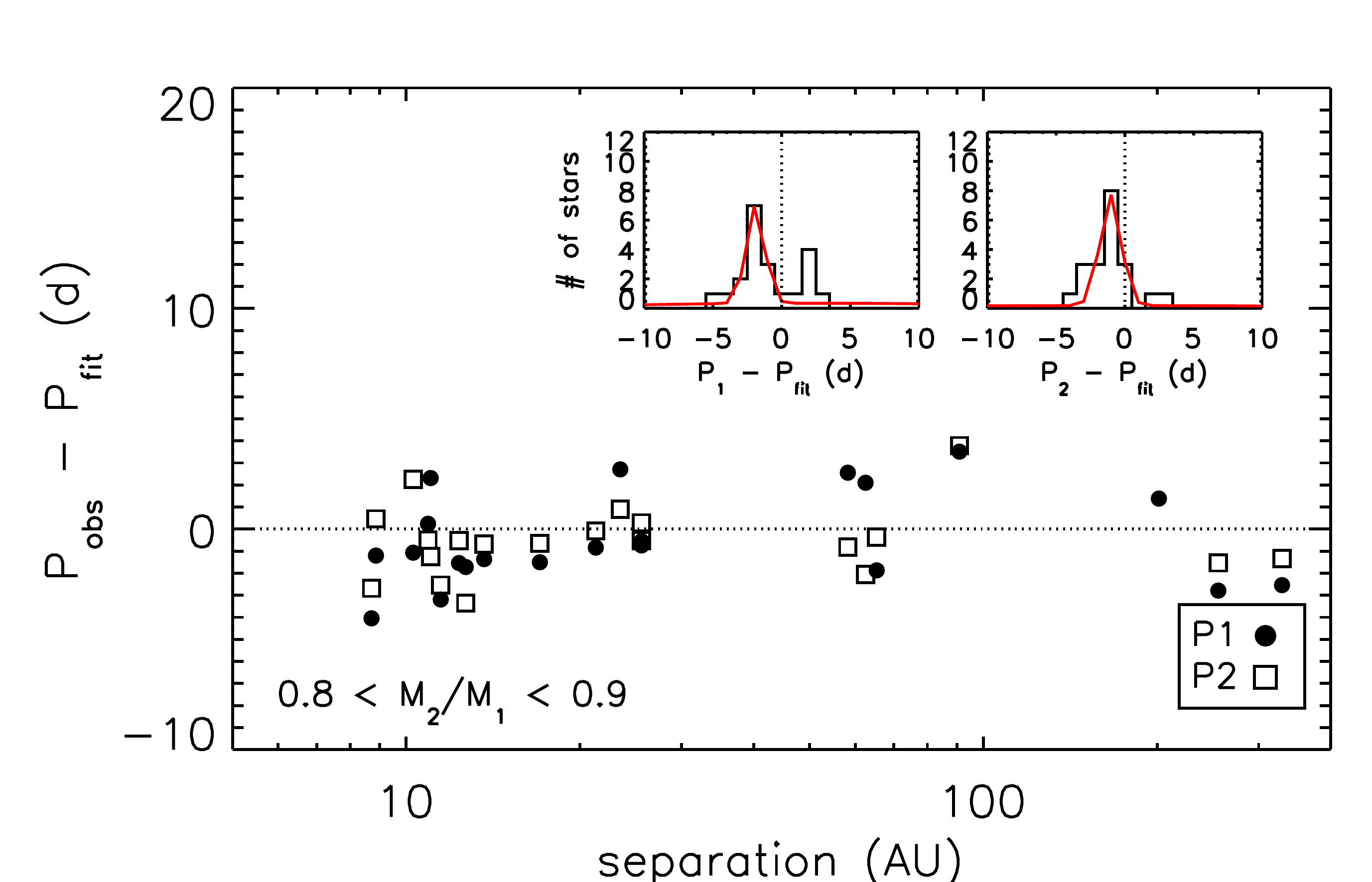}
\end{minipage}
\caption{\label{residuals1} Distribution of residuals of periods with respect to the fit (solid line in Fig.\,\ref{color-period}) for close binaries with about equal-mass components M$_2$/M$_1$ $\ge$ 0.9 (top panel) and with non-equal-mass components  0.8 $\le$ M$_2$/M$_1$ $<$ 0.9 (bottom panel). The top inner plots show the distribution of average $<$ P - P$_{fit}$ $>$ residuals from Monte Carlo simulations (see text for explanation) with a Gaussian fit over plotted; the bottom inner plots show the distribution of the  P - P$_{fit}$ residuals.}
\end{figure}
\\
\indent
We compute the difference between the primary periods P1 and the fit (solid line in Fig. \ref{color-period}) and between the secondary period P2 and the fit. This operation allows the mass dependence of the rotation period to be removed. To explore any dependence of the rotation enhancement on the mass ratio between the components of a binary system, we consider two different mass ratio ranges: binaries whose components have about equal mass  M$_2$/M$_1$ $\ge$ 0.9 and binaries with smaller mass ratios 0.8 $\le$ M$_2$/M$_1$ $<$ 0.9. 
The rotation period residuals P$_{\rm obs}$ $-$ P$_{fit}$ are plotted versus the projected separation of the components in Fig.\,\ref{residuals1}. As previously done for the single stars, components of binaries with period residuals larger than 3$\sigma$ are also excluded from the following analysis (crossed symbols in Fig.\,\ref{color-period}).  The fact that the  component of one binary has an outlying period does not necessarily imply that the binary is not a member of the USco association. For example, the spot pattern on that component may have lead to measurement of the beat period instead of the rotation period. Therefore, we only excluded
the outlying component from the analysis while keeping the rotation period of the other component. 

In the case of close binaries with components of about equal-mass,  as explained above,  we do not know to which components the P1 and P2 rotation periods refer.  Therefore, we decided to make   Monte Carlo simulations where the P1 and P2 periods of each binary are randomly permuted. We made 1000 such simulations and for each we measured the average $<$ P1 - P$_{fit}$ $>$ and $<$ P2 - P$_{fit}$ $>$. 
 The results of our simulations  are plotted in the   inner plots of the top panel of Fig.\,\ref{residuals1}, where we plot in the form of histograms the distribution of the average $<$ P - P$_{fit}$ $>$ for each of the 1000 simulations.  
We find that both periods P1 and P2 are shorter by $\simeq$ 0.4\,d with respect to the average periods of single stars. \rm
We also find that the residuals  P1$-$fit and P2$-$fit  show some marginal evidence to be correlated \rm to the projected separation between the binary components (top panel of Fig.\,\ref{residuals1}).
 The Spearman's rank correlation analysis gives   similar correlation coefficient $\rho$ = +0.29 and significance p-value  = 0.14 for the P1$-$fit and the P2$-$fit. \\ \rm
\indent
In the case of non-equal-mass components (bottom panel of Fig.\,\ref{residuals1}),   
 we find that 
  both periods P1 and P2 are shorter by $\simeq$ 1.9\,d and $\simeq$1.\,d, respectively, compared to the average period of single stars.  
The Spearman's rank correlation analysis gives a  correlation coefficient $\rho$ = +0.17  and a significance p-value  = 0.43 for P1$-$fit and  $\rho$ = -0.11  and a significance p-value  = 0.63 for P2$-$fit \rm between residuals and projected separation.

\begin{figure}
\begin{minipage}{10cm}
\includegraphics[scale = 0.32, trim = 0 0 0 0, clip, angle=90]{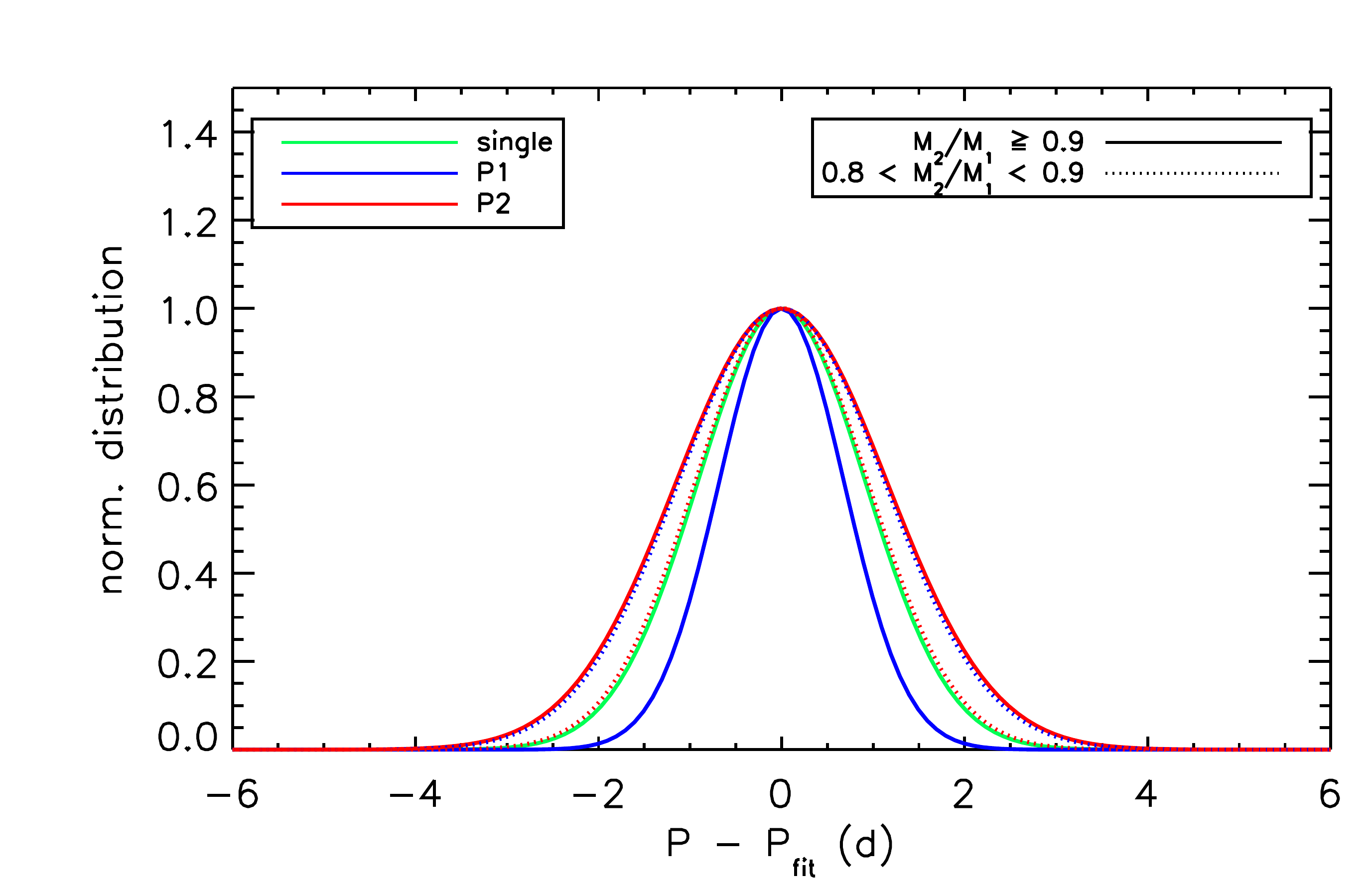}
\end{minipage}
\caption{\label{residuals} Gaussian fits to the distribution of the residuals of the rotation periods: green line for single stars, blue line  for primary components,  and red line for secondary components of binary systems.   We note that the distributions for the P1 and P2 of binary components have been shifted to be centered on zero to make the width difference more easily readable. }
\end{figure}

To summarise, we find that candidate members of USco in close binaries (median separation $\sim$ 21\,AU in the analysed sample) rotate faster than their single counterparts; moreover,  the lower-mass components of non-equal-mass binaries tend to  have the shortest rotation periods.  \rm We find some marginal evidence that the closer the equal-mass binary components, the faster their rotation period with respect to single stars.\\
\indent
Another property of our sample of close binaries, in addition to the average rotation period, is the period difference between the two components. 
 We have seen that the rotation period is mass dependent; therefore, the period difference between the two components of the same system may arise on only their mass difference. For this reason, we first remove the mass dependence by computing the residuals with respect to a linear fit to the periods before computing the period differences. We find that the average period difference between components of approximately equal-mass binaries (M$_2$/M$_1$ $\ge$ 0.9)  is $<$$\Delta$P$> $ $\simeq$ 0.8\,d (with a dispersion $\sigma$ = 1.6\,d) against  $<$$\Delta$P$>$  $\simeq$ 0.2\,d (with a dispersion $\sigma$ = 1.9\,d) between the components of unequal-mass binaries (0.8 $\le$ M$_2$/M$_1$ $<$ 0.9).   \\ \rm
\indent
The last property that we take into consideration is the width of the period distribution (see Fig.\,\ref{residuals}). After removing the mass dependence, as already done before,    we find that among binaries with approximately equal-mass components, primary  and secondary components have residual distribution, respectively, smaller ($\sigma$ $\simeq$ 0.7\,d) and larger ($\sigma$ $\simeq$ 1.1\,d) than single stars ($\sigma$ $\simeq$ 0.9\,d). Alternatively, among binaries with non-equal-mass components,  secondaries   have similar residual distribution to single stars ($\sigma$ $\simeq$ 0.9\,d), whereas primaries have a larger width of the residual distribution with  a standard deviation $\sigma$ $\simeq$ 1.1\,d.

Another important aspect concerns the fraction of discs.
In the colour range under analysis, about 28\% of candidate members that are single (i.e. with only one periodicity measured) show strong evidence for a disc. If we consider the resolved close binaries in the same colour range, we find that the fraction decreases to 14\%.

\begin{figure*}[!h]
\begin{minipage}{18cm}
\includegraphics[scale = 0.52, trim = 0 0 0 0, clip, angle=90]{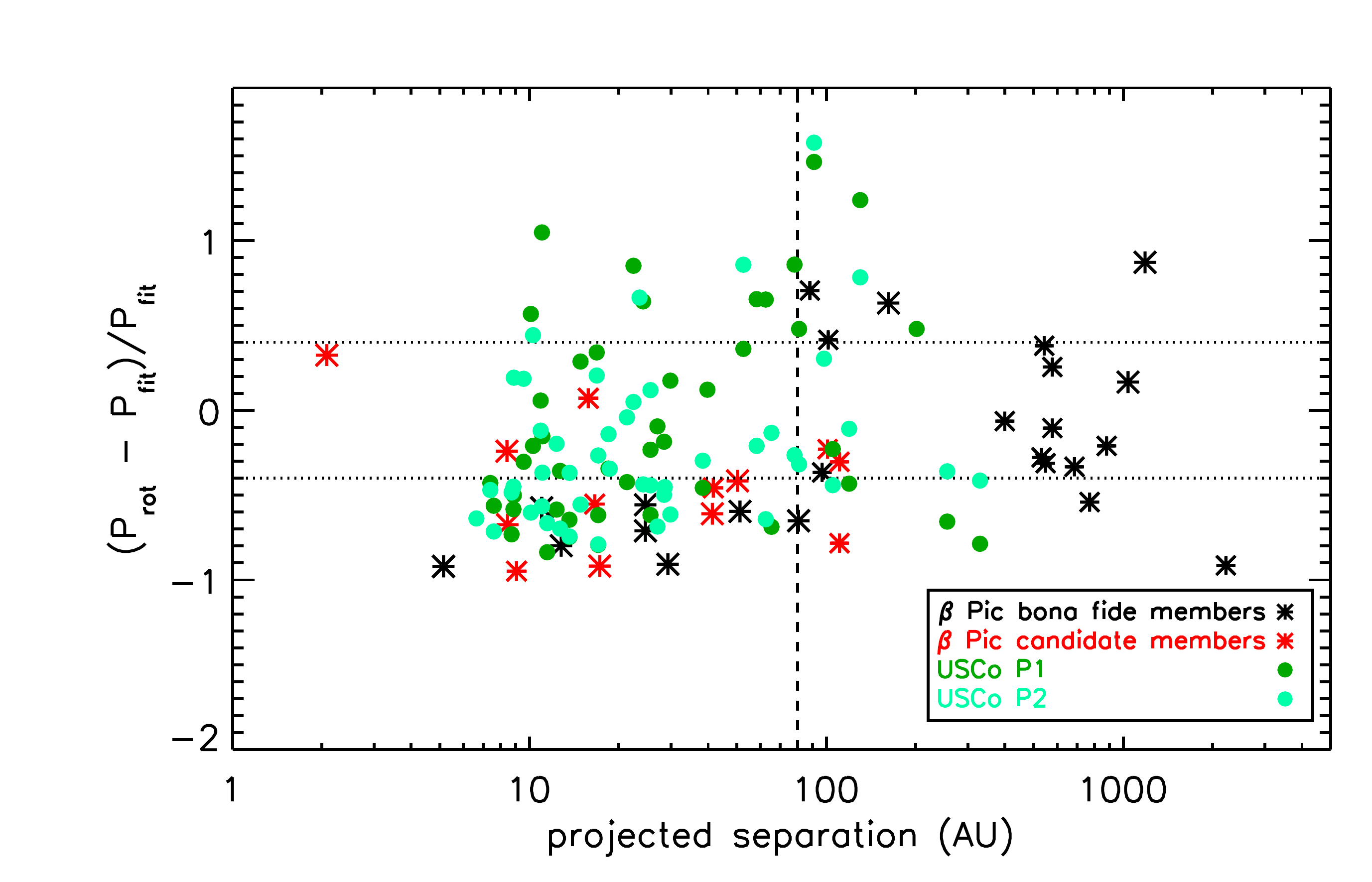}
\end{minipage}
\caption{\label{comparison} Comparison between the distribution of relative period residuals vs. projected separation for the members of the $\beta$ Pic association (see also Fig.\,7 of \citealt{Messina17}) and the close-binary candidate members of Upper Scorpius. Horizontal dotted lines represent the width ($\pm$3$\sigma$) of the period distribution of the $\beta$ Pic single members. In the  $\beta$ Pic association, close binaries with component separation $<$80 AU (vertical dashed line)  mostly rotate faster than single stars.}
\end{figure*}

\section{Discussion}

The disc lifetime is variable, but generally not longer than $\sim$10\,Myr (\citealt{Ingleby14}, \citealt{Ribas14}). Theories supported by observations predict that the disc lifetime can be significantly shortened by the presence of a companion (\citealt{Meibom07}, \citealt{Bouvier93}, \citealt{Edwards93}, \citealt{Ingleby14}, \citealt{Rebull04}).
\citet{Kraus16} and \citet{Cieza09} found that stars without IR excess tend to have companions at smaller separation than 
stars with excess indicating the presence of a disc. Both studies find that the depletion of primordial discs among binary systems with 
components closer than 40 AU is a factor of two larger than in either single or wide binaries already at ages as young as 1-2 Myr.  \citet{Stauffer16}  report that photometric binaries among the Pleiades GKM-type
stars tend to rotate faster than their counterpart single stars.
\citet{Douglas16}  report that most, if not all, rapid rotators that deviate from the single-valued relation between mass and  rotation already reached by the age of the Hyades ($\sim$0.6 Gyr), belong to multiple systems. \\
Recently, \cite{Messina17}, analysing the rotation period distribution of the members of the 25-Myr \object{beta Pic Association,}  found that single stars and components of multiple systems with projected separation larger than about 80\,AU  have similar distribution of rotation periods versus V$-$K$_s$ colour. 
On the contrary, components of close visual binaries/triplets with projected separation smaller than about 80\,AU  rotate preferentially faster than their equal-mass single counterparts. This circumstance suggests that when the components are sufficiently close,  their primordial discs undergo an enhanced dispersal allowing the stars to start their spin-up earlier than single stars. \\
The results by  \cite{Messina17} for stars of 25 Myr of age can be compared to those we found for stars of 8 Myr for the USco association. In Fig.\ref{comparison}, we plot the results of \cite{Messina17}, that is the relative residuals of the fit to the rotation period versus the projected separation (AU) as asterisks and overplot the same quantity but for the resolved binaries considered in this study as bullets. The use of residuals allows us to remove the mass dependence in the period distribution.
We note that in the 5--80 AU range of projected separation about 70\%  of close binaries  at the younger age of USco have periods shorter (negative residuals) than their single counterparts, and about 30\% still have periods comparable to those of the single stars. Conversely, at the older age of  the $\beta$ Pic association, all close binaries have periods shorter than their single counterparts.
This is a clear indication that the post-disc dispersal stellar rotation spin up is already set at  ages younger than 8 Myr, and that it has produced measurable effects on the majority of close binaries by an age of 8 Myr. The disc dispersal timescale in these close binaries  must be different from binary to binary, with a range of values, allowing for close binaries  (about 30\% in this sample) where the dispersal takes place slowly, making the rotation spin-up not yet effective, as well as binaries where the dispersal was quite sudden making the rotation spin-up measurable.\\
We also note that the rotation-period shortening for components of equal-mass binaries ($\sim$0.4\,d for P1 and P2) is smaller than for non-equal-mass binaries  ($\sim$1.9\,d for P1 and $\sim$1.0d for P2). This suggests that the timescale of their disc dispersal is  dependent  on the mass ratio between the binary components.\\
 
 \section{Conclusions}
 We analysed a sample of 49 close binaries that are candidate members of the Upper Scorpius association whose components are in the M0--M6 spectral range, and have known rotation periods and projected separations ($\rho <$ 100\,AU).
We found clear evidence that they rotate faster than their single counterparts. On average, components of close binaries exhibit rotation periods shorter by an amount ranging from 0.4\,d if they have about equal-mass, to 1.9\,d as in the case of the lower-mass components of lower mass-ratio binaries.
 The rotation spin up of close binaries with respect to single stars can be attributed, among different processes, to an early dispersion or truncation of the primordial circumstellar disc owing to gravitational effects by the close companion. Such a disc dispersal likely starts operating in the very first few million years of stellar life, producing measurable effects at the age of 8 Myr.\\
In our hypothesis that the rotation period shortening with respect to single stars is a direct consequence of early disc dispersal, we infer that the timescale of disc dispersal is the longest in single stars or in wide-orbit ($\rho$ $\ga$ 100\,AU) components of multiple systems, is  shorter in binaries with about equal-mass components, and is even shorter in binaries with non-equal-mass components. \rm
Finally, we find that components of about equal-mass and of non-equal-mass binaries generally have different rotation period dispersion widths, where primary components of equal-mass binaries exhibit a smaller dispersion than that of single stars, whereas secondaries and components of non-equal-mass binaries all exhibit dispersion comparable to or larger than that of single stars. 
\\

 \begin{table*}
 \caption{\label{Table1} EPIC ID number of the 49 binaries considered in the present study; colour-corrected integrated (V$-$K)$_0$ colour,  colours, masses, and rotation periods for the primary and secondary components, respectively; I magnitude of the whole system, and magnitude difference between the two components.}
          \begin{tabular}{ccccccccccr}
          \hline
          EPIC & (V$-$K)$_0$$^{(a)}$ & (V-K)$_{0P}$ & (V-K)$_{0S}$ & M1$^{(b)}$ & M2 & P1$^{(a)}$ & P2$^{(a)}$ &  I mag$^{(b)}$ & $\Delta$I$^{(b)}$ & separation$^{(b)}$\\
          ID & (mag) & (mag) & (mag) & (M$_\odot$) & (M$_\odot$)   &(d) & (d) & (mag) & (mag) & (AU)\,\,\,\,\,\,\,\,\, \\
          \hline
     \object{EPIC 203553934}   & 4.65   &  4.61   &  4.69   &  0.59   &  0.55   &  4.70   &  6.22   & 11.47   &  0.20   &  52.57 \\
      \object{EPIC 204918279}   & 6.23   &  6.12   &  6.38   &  0.29   &  0.25   &  0.46   &  0.47   & 12.60   &  0.20   &  25.57 \\
      \object{EPIC 204104740}   & 4.83   &  4.81   &  4.85   &  0.56   &  0.54   &  5.57   &  3.06   & 11.55   &  0.10   &  22.41 \\
      \object{EPIC 204832936}   & 4.98   &  4.87   &  6.20   &  0.38   &  0.03   &  4.26   &  4.98   & 12.86   &  1.80   & 201.46 \\
      \object{EPIC 204477741}   & 5.86   &  5.86   &  5.86   &  0.31   &  0.31   &  0.82   &  0.76   & 12.74   &  0.00   &   7.39 \\
      \object{EPIC 204878974}   & 4.22   &  4.14   &  4.33   &  0.56   &  0.48   &  3.09   &  0.85   & 11.16   &  0.40   &  12.68 \\
      \object{EPIC 204350593}   & 5.13   &  5.10   &  5.17   &  0.42   &  0.40   &  3.16   &  0.98   & 12.16   &  0.10   &  14.87 \\
      \object{EPIC 204406748}   & 4.52   &  4.43   &  4.64   &  0.69   &  0.57   &  4.39   & 16.96   & 11.14   &  0.60   &  39.78 \\
      \object{EPIC 203690414}   & 5.33   &  5.15   &  5.65   &  0.32   &  0.24   &  1.82   &  1.95   & 12.95   &  0.40   &  25.57 \\
      \object{EPIC 204794876}   & 4.02   &  3.94   &  4.13   &  0.73   &  0.61   &  1.49   &  2.15   & 10.87   &  0.60   &   8.71 \\
      \object{EPIC 204862109}   & 4.15   &  4.12   &  4.18   &  0.73   &  0.69   &  1.73   &  1.05   & 10.88   &  0.20   &  13.64 \\
      \object{EPIC 204637622}   & 5.06   &  5.06   &  5.06   &  0.50   &  0.50   &  1.05   &  1.39   & 11.99   &  0.00   &   8.84 \\
      \object{EPIC 204603210}   & 5.19   &  5.00   &  5.59   &  0.39   &  0.27   &  1.09   &  1.22   & 12.42   &  0.60   &  12.35 \\
      \object{EPIC 204856827}   & 5.15   &  5.15   &  5.15   &  0.44   &  0.44   &  1.65   &  2.81   & 12.13   &  0.00   &   9.56 \\
     \object{EPIC  204242152}   & 5.08   &  5.08   &  5.08   &  0.45   &  0.45   &  2.10   &  1.57   & 12.33   &  0.00   &  11.06 \\
     \object{EPIC  204757338}   & 5.75   &  5.75   &  5.75   &  0.30   &  0.30   &  2.30   &  1.06   & 12.94   &  0.00   &  80.98 \\
     \object{EPIC  204845955}   & 5.22   &  5.13   &  5.34   &  0.53   &  0.45   &  1.20   &  2.53   & 11.66   &  0.40   &   8.86 \\
      \object{EPIC 204229583}   & 4.89   &  4.71   &  5.32   &  0.55   &  0.35   &  0.69   &  0.77   & 11.54   &  1.00   & 329.27 \\
      \object{EPIC 203851147}   & 5.33   &  5.13   &  5.75   &  0.38   &  0.26   &  5.91   &  5.34   & 12.37   &  0.60   &  90.92 \\
      \object{EPIC 205087483}   & 4.98   &  4.98   &  4.98   &  0.46   &  0.46   &  1.51   &  2.38   & 12.23   &  0.00   & 119.34 \\
      \object{EPIC 205177770}   & 5.65   &  5.40   &  6.35   &  0.30   &  0.18   &  1.14   &  0.93   & 12.91   &  0.60   &  21.31 \\
      \object{EPIC 204429883}   & 5.74   &  5.74   &  5.74   &  0.38   &  0.38   &  0.32   &  0.33   & 12.20   &  0.00   &  17.05 \\
      \object{EPIC 204608292}   & 4.50   &  4.35   &  5.97   &  0.36   &  0.10   &  4.40   &  0.83   & 12.38   &  1.30   &  10.91 \\
      \object{EPIC 203036995}   & 5.33   &  5.11   &  5.87   &  0.38   &  0.24   &  0.93   &  0.78   & 12.39   &  0.70   &  17.05 \\
      \object{EPIC 203716047}  & 5.14   &  4.95   &  5.54   &  0.39   &  0.27   &  0.85   &  1.44   & 12.43   &  0.60   &  65.35 \\
      \object{EPIC 204857023}   & 4.44   &  4.31   &  4.64   &  0.54   &  0.42   &  1.46   &  1.86   & 12.46   &  0.60   & 255.36 \\
      \object{EPIC 203048597}   & 5.33   &  5.33   &  5.33   &  0.39   &  0.39   &  1.61   &  1.17   & 12.31   &  0.00   & 105.13 \\
      \object{EPIC 204156820}   & 4.57   &  4.57   &  4.57   &  0.59   &  0.59   &  6.61   &  2.62   & 11.74   &  0.00   &  78.14 \\
      \object{EPIC 205225696}   & 5.19   &  5.19   &  5.19   &  0.34   &  0.34   &  2.70   &  0.89   & 13.26   &  0.00   &  29.83 \\
      \object{EPIC 203777800}   & 4.95   &  4.91   &  4.99   &  0.39   &  0.37   &  2.28   &  1.25   & 12.65   &  0.10   &  28.41 \\
      \object{EPIC 204449800}   & 6.25   &  6.25   &  6.25   &  0.31   &  0.31   &  2.45   &  1.95   & 12.58   &  0.00   & 130.05 \\
      \object{EPIC 204235325}   & 5.11   &  4.88   &  6.19   &  0.44   &  0.20   & 24.03   &  2.01   & 12.11   &  1.20   &  98.19 \\
      \object{EPIC 202533810}   & 4.78   &  4.78   &  4.78   &  0.52   &  0.52   &  5.04   &  1.73   & 11.80   &  0.00   &  24.15 \\
      \object{EPIC 204204606}   & 5.31   &  5.31   &  5.31   &  0.33   &  0.33   &  1.15   &  1.49   & 12.86   &  0.00   &  38.36 \\
      \object{EPIC 204082531}   & 5.81   &  5.81   &  5.81   &  0.27   &  0.27   &  0.65   &  0.43   & 12.78   &  0.00   &   7.59 \\
      \object{EPIC 203855509}   & 5.41   &  5.25   &  5.68   &  0.35   &  0.27   &  4.52   &  0.39   & 12.93   &  0.40   &  11.03 \\
      \object{EPIC 203809317}   & 4.55   &  4.44   &  4.72   &  0.52   &  0.42   &  6.46   &  2.39   & 12.36   &  0.50   &  58.25 \\
      \object{EPIC 204569229}   & 4.22   &  4.16   &  4.30   &  0.66   &  0.58   & 22.92   &  1.26   & 11.27   &  0.40   &   6.63 \\
      \object{EPIC 204655550}   & 4.20   &  4.06   &  4.47   &  0.63   &  0.45   &  4.03   &  6.08   & 11.41   &  0.90   &  10.28 \\
      \object{EPIC 204374147}   & 4.59   &  4.47   &  4.80   &  0.71   &  0.53   &  0.62   &  0.49   & 11.05   &  0.90   &  11.48 \\
      \object{EPIC 203001867}   & 4.78   &  4.71   &  4.86   &  0.53   &  0.47   &  5.32   &  0.83   & 11.91   &  0.30   &  62.51 \\
     \object{EPIC 204462113}   & 4.60   &  4.53   &  4.69   &  0.40   &  0.36   &  5.73   &  1.07   & 12.69   &  0.20   &  10.12 \\
      \object{EPIC 203071614}   & 5.72   &  5.52   &  6.11   &  0.29   &  0.21   &  0.46   &  0.53   & 12.55   &  0.40   &  13.65 \\
      \object{EPIC 204520585}   & 4.25   &  4.13   &  4.46   &  0.70   &  0.52   & 19.16   &  2.16   & 11.75   &  0.90   &  18.65 \\
      \object{EPIC 203856244}   & 6.08   &  5.94   &  6.31   &  0.44   &  0.34   &  4.07   &  1.95   & 12.11   &  0.50   &  23.49 \\
      \object{EPIC 203850605}   & 4.85   &  4.85   &  4.85   &  0.60   &  0.60   &  3.92   &  3.52   & 10.76   &  0.00   &  16.86 \\
      \object{EPIC 203115615}   & 4.73   &  4.58   &  5.02   &  0.65   &  0.45   & 26.17   &  0.99   & 12.36   &  1.00   &  28.58 \\
      \object{EPIC 202615424}   & 4.30   &  4.30   &  4.30   &  0.52   &  0.52   &  2.82   &  3.69   & 12.34   &  0.00   &  18.47 \\
      \object{EPIC 203873374}   & 5.23   &  5.20   &  5.26   &  0.48   &  0.46   &  2.06   &  0.63   & 12.50   &  0.10   &  26.99 \\
      \hline
      \multicolumn{10}{l}{(a): values taken from \citealt{Rebull18}; (b): values taken from \citealt{Tokovinin18}}\\
 \end{tabular}
 \end{table*}
 {\it Acknowledgements}. Research on stellar activity at INAF- Catania Astrophysical Observatory 
is supported by MIUR  (Ministero dell'Istruzione, dell'Universit\`a e della Ricerca).  This research has made use of the Simbad database operated at CDS (Strasbourg, France).   We are grateful to the referee who allowed us to significantly improve the quality of this paper.
\\

\bibliographystyle{aa.bst} 
\bibliography{usco} 

\end{document}